\documentclass{vgtc}                          

\ifpdf
  \pdfoutput=1\relax                   
  \usepackage{graphicx}                
  \DeclareGraphicsExtensions{.pdf,.png,.jpg,.jpeg} 
\else
  \usepackage{graphicx}                
  \DeclareGraphicsExtensions{.eps}     
\fi%

\graphicspath{{figures/}{pictures/}{images/}{./}} 
\usepackage{caption}

\usepackage{microtype}                 
\PassOptionsToPackage{warn}{textcomp}  
\usepackage{textcomp}                  
\usepackage{mathptmx}                  
\usepackage{times}                     
\usepackage{cite}                      
\usepackage{tabu}                      
\usepackage{booktabs}                  

\onlineid{W8-iXR-1}

\vgtccategory{Research}

\vgtcinsertpkg

\title{Enabling Data-Driven and Empathetic Interactions: A Context-Aware 3D Virtual Agent in Mixed Reality for Enhanced Financial Customer Experience}
\author{Cindy Xu\thanks{e-mail: cindy.x.xu@jpmorgan.com} %
\and Mengyu Chen\thanks{e-mail:mengyu.chen@jpmchase.com} %
\and Pranav Deshpande \thanks{e-mail:pranav.deshpande@jpmchase.com} %
\and Elvir Azanli\thanks{e-mail:elvir.azanli@jpmorgan.com} %
\and Runqing Yang\thanks{email:runqing.yang@jpmorgan.com}
\and Joseph Ligman\thanks{e-mail:joseph.ligman@jpmchase.com}}
\affiliation{\scriptsize Global Technology Applied Research, JPMorganChase}

\teaser{
 \centering
 \includegraphics[width=0.8\linewidth, trim={0 26 0 17},clip]{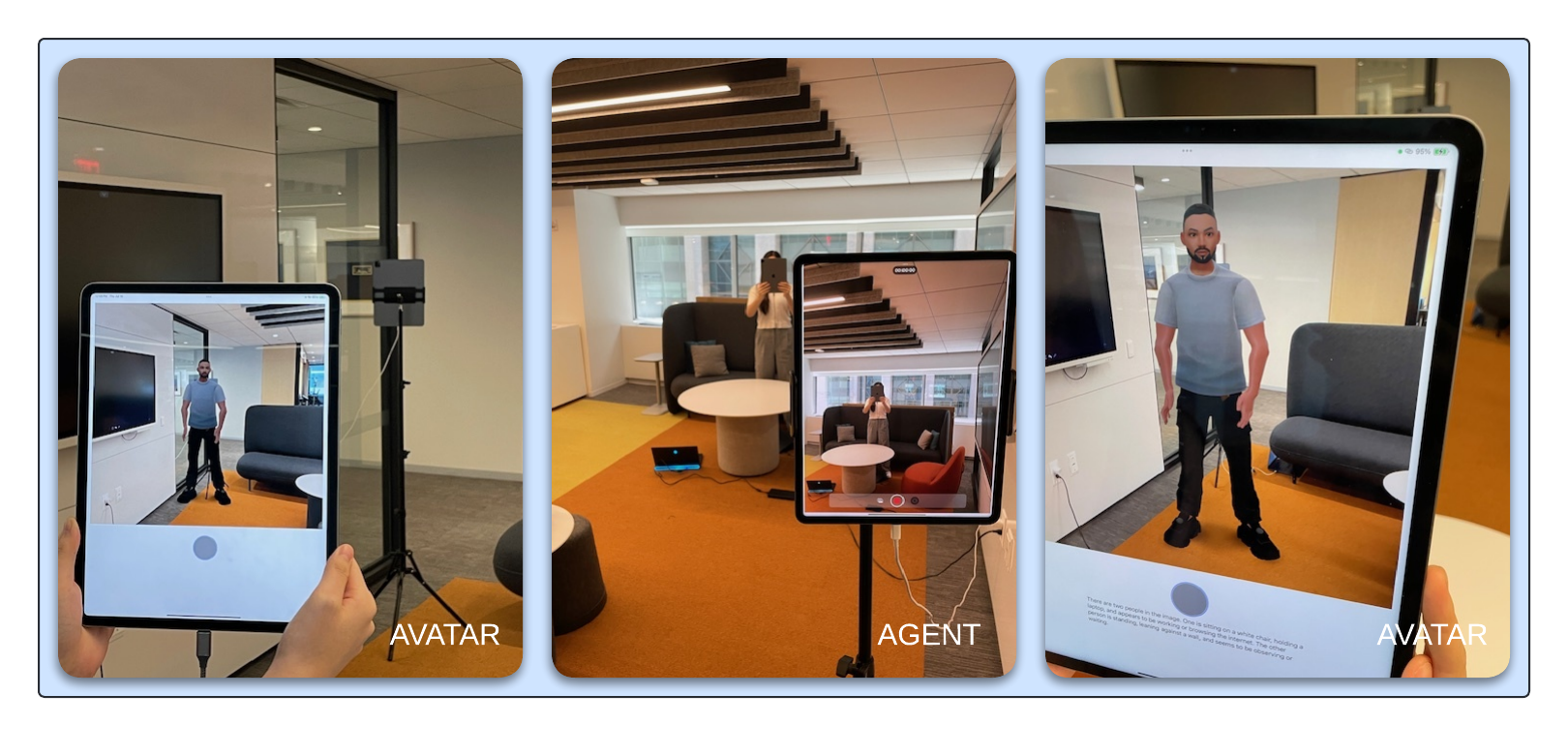}
 \caption{A context-aware 3D virtual agent in mixed reality. Left: User POV with interactive chatbot interface. Middle: Agent POV displaying field-of-view and video camera input. Right: Close-up of the embodied agent projected in AR.}
 \label{fig:teaser}
}

\abstract{
In this paper, we introduce a novel system designed to enhance customer service in the financial and retail sectors through a context-aware 3D virtual agent, utilizing Mixed Reality (MR) and Vision Language Models (VLMs). Our approach focuses on enabling data-driven and empathetic interactions that ensure customer satisfaction by introducing situational awareness of the physical location, personalized interactions based on customer profiles, and rigorous privacy and security standards. We discuss our design considerations critical for deployment in real-world customer service environments, addressing challenges in user data management and sensitive information handling. We also outline the system architecture and key features unique to banking and retail environments. Our work demonstrates the potential of integrating MR and VLMs in service industries, offering practical insights in customer service delivery while maintaining high standards of security and personalization.
} 

\CCScatlist{ 
  \CCScat{H.5.1}{INFORMATION INTERFACES AND PRESENTATION (e.g., HCI)}%
{Multimedia Information Systems}{Artificial, augmented, and virtual realities};
  \CCScat{K.4.3}{COMPUTERS AND SOCIETY}{Organizational Impacts}{Computer-supported collaborative work}
}

\begin{document}


\firstsection{Introduction}

\maketitle

Navigating the complexities of customer service in finance and retail environments presents a series of challenges that can significantly impact customer satisfaction and operational efficiency. Common issues such as prolonged wait times, insufficient service agents (e.g., bank tellers, store clerks, sales associates), and the inconvenience of multiple visits for minor inquiries or transactions are frequent pain points for customers. For instance, in a typical banking scenario, customers might endure lengthy delays due to a limited number of bank tellers available to manage a wide range of transactional needs. Effective in-branch customer service requires efficient staff scheduling for adequate coverage during peak hours and a well-organized and accessible interior layout to enhance the overall customer experience \cite{tlapana2021}. Additionally, branch managers need to provide thorough training for service agents to equip them with the necessary service knowledge and customer handling skills. Challenges often arise in balancing operational costs with the need for high-quality service, and integrating technology or automated processes to streamline operations while maintaining a personal touch in customer engagement.

To improve customer experience in banking and retail environments, various methods have been explored, such as interactive kiosks \cite{Lao2021ExploringHD}, digital queuing systems \cite{obermeier2020}, and online or mobile services \cite{guo2023mobile}. These technologies are primarily designed to reduce human involvement in routine tasks, aiming to streamline operations and minimize wait times. However, lacking the support by a professional service agent, still, these methods are typically self-driven by customers and do not proactively offer personalized assistance. This passive nature limits their effectiveness in increasing customer satisfaction and prevents them from further providing tailored and accurate service that meet the unique needs of each customer. 

In light of these challenges, we are interested in making in-branch customer service more dynamic and intelligent, so that customers may be served more seamlessly, relying less on their own navigation of services and gaining more proactive support that directly addresses their personal needs and concerns. Vision Language Models (VLMs) have huge opportunities as they could provide extensive scene understanding and intelligence for the entire customer situation, including both individual-level insights and comprehensive store-level analysis. This situational knowledge can help us design more effective client service delivery and interaction models. In this paper, we explore the design considerations of a MR-based intelligent 3D virtual agent in the context of financial services and retail-store customer experience. We discuss our early system implementation that utilizes multiple mobile cameras with a Mixed Reality (MR) headset to acquire multi-view images as VLM prompts. By leveraging the situational knowledge about a customer and natural interaction, our 3D virtual agents in MR can interpret the visual context and anticipate customer needs, offering personalized and relevant solutions in real-time.

\section{Prior Work}
\subsection{Vision Language Models in Virtual Agents}
 Following the emergence of Large Language Models (LLMs), the most recent development in artificial intelligence explores the modality of vision. Vision Language Models (VLMs) are multi-modal models trained on images and language, designed to understand the relationship between image and text inputs and generate human-like responses. VLMs bridge the gap between natural language processing and computer vision capabilities, enabling advanced visual understanding tasks such as visual question answering\cite{chen2022pali, beit3}, visual summarization\cite{li2022mplug, wang2022git}, and image retrieval \cite{bordes2024introduction}. Recently, using LLMs to design virtual agents with AI chatbot capabilities has been a greatly explored topic \cite{guo2024largelanguagemodelbased},\cite{wan2024},\cite{zhang2024buildingcooperativeembodiedagents}. However, VLMs pose a significant development, opening up avenues for designing intelligent, multi-modal agents with both vision and language understanding. Recent works propose different applications for VLM-powered agents, such as instruction-following agents \cite{liu2023instructionfollowing}, on-screen assistants for GUI navigation tasks \cite{niu2024screenagentvisionlanguagemodeldriven}, and intelligent virtual reality systems  \cite{konenkov2024vrgptvisuallanguagemodel}. In our work, we are interested in embedding VLMs in our virtual agent system in order to develop an intelligent assistant with real-time sight, capable of using its contextual awareness of the surrounding physical environment to engage with users in a more proactive and meaningful way.

\subsection{Embodied Conversational Agents}
Embodied conversational agents (ECAs), more commonly known as "chatbots" \cite{schobel2023}, are digital 2D or 3D graphic representations of AI agents that utilize Natural Language Processing (NLP) to engage in dialog with users and provide assistance in response to user inquiries \cite{bailenson2004}. Past research into ECAs, particularly in virtual environments, has emphasized the critical role of the visual appearance of the agent in improving user experience and interaction. Crolic et al. describes how in chatbot-led service interactions, users respond more positively to anthropomorphic avatars that resemble human features and behaviors, which helps establish customer trust \cite{crolic2021}. Designing an embodied agent with a human-like appearance and gestures thus becomes crucial for applications such as customer service and retail assistance, where effective communication and user comfort are essential. Currently, state-of-the-art ECAs rely mainly on LLMs, such as "FurChat," an advanced conversational AI robot receptionist built on GPT-3.5 \cite{cherakara2023}. However, there is little to no work exploring how vision language models can be embedded within ECAs in order to imbue an agent with visual understanding as well as natural language processing. By integrating VLMs, our research aims to further explore the capabilities of ECAs, allowing an agent to not only understand and respond to natural language but also perceive and react to visual cues in mixed reality scenarios.

\subsection{Augmented Reality and Interactive Agents}
Augmented reality (AR) integrates digital information with a user's environment in real-time, enhancing their perception of the physical world through overlaid visual imagery. In their 1998 paper, Nagao laid the foundation for interactive AR agents with situational awareness, defining the "real world agent" as a special agent capable of recognizing users' real world situations and supporting tasks as needed \cite{Nagao1998AgentAR}. One proposed application is a non-embodied shopping assistant agent for navigational assistance in stores and commercial object recognition. While their system is built using electromagnetic sensors and infrared ID (location beacons), our work builds upon their work, utilizing novel technologies such as VLMs and IoT sensors for location tracking. Recent work has also explored user interaction with virtual AI agents within AR environments, focusing on studying user engagement and agent design. Wang et al. recently compared user perception and satisfaction of different virtual agent representations for AR headsets, a key finding being that users greatly prefer embodied agents over voice-only assistants \cite{wang2019}. For customer service and retail applications, mixing the digital and real world spaces provides opportunity to enhance online and in-store retail \cite{bonetti2018}, B2B customer experience \cite{wieland2024} as well as the financial and banking sectors as proposed.
\section{Design Considerations}
There are several end-user needs and requirements we are considering: 1) proactive and personalized support, 2) situational awareness, and 3) privacy assurance.

\subsection{Proactive and Personalized Support}
Customers often visit banks and retail stores with time-sensitive needs. With many services available online these days, the reasons for a customer to visit a physical location are frequently urgent or complex. Therefore, reducing wait times and efficiently identifying unique situations may be prioritized to help determine the most effective way or resource to support them. Existing strategies such as self-service options like kiosks or mobile apps often require customers to independently identify solutions to their issues. While this can be efficient for straightforward tasks, it becomes less optimal for more complex problems. In such cases, customers may struggle to locate the appropriate resource or entry point, potentially leading to further delays, increased wait times, or even wasted effort before their issues can be effectively addressed. This highlights the need for a more guided and supportive approach, particularly for intricate or urgent customer concerns, where direct and proactive assistance can significantly enhance the resolution process.

\subsection{Situational Awareness}
In a bank branch or a retail store, dynamic situations frequently arise that may impact the customer experience. These situations include high customer traffic, insufficient guidance or unclear instructions, and unexpected events such as system outages or emergencies. All of these scenarios can lead to customer frustration, causing confusion, delay, or a perceived lack of support. Addressing these dynamic challenges is crucial for maintaining a high level of customer satisfaction and operational efficiency. Advanced situational awareness capabilities for an AI agent are highly desired as it allows for real-time monitoring and analysis of the environment, such as tracking customer density in different areas, identifying or predicting peak times, and detecting unusual activities. With such knowledge, both human staff and the virtual agent may work together to implement optimization strategies to adjust store operations and manage resources more efficiently.

\subsection{Privacy Assurance}
Privacy assurance is a critical component in the design of any technology that handles sensitive information, particularly in the financial sector where customer trust is a top priority. When designing a context-aware 3D virtual agent, it is essential to ensure robust privacy and security measures are in place to protect customer related data such as profiles, account information, and transaction data. Enterprise-level encryption methods should be employed for data at rest and in transit, alongside strict access controls and regular security inspections to ensure that only authorized personnel and system components have access to sensitive data. Especially for a context-aware 3D virtual agent, multi-modal interaction such as vision and speech often requires the usage of different system modules (e.g., speech recognition modules, machine learning models) to process customer input, and such input may contain sensitive information that needs to be securely handled. Therefore, each component that handles customer data must be designed with robust security measures to prevent authorized access or data breaches. End-to-end encryption for data transmission, secure data storage, and role or rule based access controls need to be considered to protect the integrity and confidentiality of the data. At the same time, customers should also be informed about the data processing activities performed by the virtual agent and given control over their personal information, including the ability to opt-out of certain data collection if desired. Transparency in how technology is used to respect their privacy is crucial for maintaining a satisfactory customer experience.

In addition to the enterprise-level privacy and security measures outlined above, an emphasis on usable privacy from the end-user perspective should also be considered, with a particular focus on user \textit{awareness} of privacy risks, \textit{motivation} to protect themselves, and \textit{ability} to act within the system capabilities\cite{das2023}. While the system should be well designed by the service provider, it should also empower customers with the knowledge and tools necessary to actively participate in the protection of their privacy.

\section{System Design}
Our goal is to support a seamlessly automated process for banking and in-store retail customers while providing a highly personalized virtual agent experience to address individual needs by both visual and language intelligence. Building on the design considerations above, our virtual agent system supports proactive interactions for personalized assistance based on both explicitly expressed needs and implicitly revealed visual context by the customers, as well as comprehensive store-level insights gained from multiple camera inputs, allowing the virtual agent to manage customer interaction at different levels. Technically, the system will let us explore different strategies in guiding customers through the banking and retail experience, such as customer localization, queue management, and personalized recommendation. Our system has been designed to support a robust 3D virtual agent with the ability to provide a data-driven execution as well as an empathetic and trustful user experience. 

The virtual agent system is specifically designed with distributed embodiment to provide the agent with access to camera views at different perspectives. The customer's mixed reality application, as a primary entry point for customer experience, can be deployed on a handheld tablet or an AR headset to let the customer see and interact with the virtual agent's digital representation projected in to the physical environment. The distributed applications work together allowing a customer to interact with an agent. The customer uses their own device, while the virtual agent is hosted on one or more in-store camera-enabled devices. This setup provides for a centralized and secure deployment. \autoref{fig:teaser} shows an example setup of the virtual agent system with this concept of distributed embodiment. 

\subsection{Proactive Virtual Agent Interaction}
We propose a system where customer service and retail settings are enhanced with intelligent agents that can directly engage and interact with customers. A mixed reality solution with a physically located agent representation in a retail setting provides unique advantages to the individual experience of online banking and the in-person retail experience. Specifically, we utilize mixed reality to improve the overall customer experience in the following ways:

\textbf{\textit{Face-to-Face Communication}} - In a retail setting, it is essential to foster customer relationships while providing a fully digitized solution. We can maintain human-like customer interaction by allowing users to engage with virtual agents, ensuring a personal and comfortable experience. We designed animations of the virtual agent based on client interactions and specific tasks (e.g., general inquiry, transaction request, information look-up), enhancing the human-like experience. By incorporating facial expressions, hand gestures, and body language that align with the context of the conversation and environmental factors, the agent can provide a more engaging and relatable interaction. This dynamic animation approach ensures that the agent not only responds verbally but also visually, mirroring human communication patterns and making the digital interaction feel more natural and intuitive. We also investigate building an agent with emotional intelligence. Using microphone audio sensors, our system is able to provide real-time feedback with analysis for emotional cues and sentiment to better assist the customer, a feature commonly used in customer service today\cite{li2019acousticsentiment, jabbar2019}.


\textbf{\textit{Dynamic and Flexible Role-Switching}} - While guiding customers through different levels of banking or retailing processes, as the dialog progresses, the virtual agent can assume different roles according to the stages of support, such as from customer service agent to financial advisor, or to sales associate. In mixed reality, it is easy to change their appearance for better engagement, and this role switching capability makes it easy to address different customer needs. Depending on business needs or focus, the bank or the store can also switch the agent roles to accommodate more specific requirements from customers. Adjusting the number of roles based on customer statistics such as peak hour visit reasons or daily staffing schedules could make the allocation of customer support resources more effective.

\textbf{\textit{Guided Multi-Location Assistance}} - Mixed reality offers the possibility to let our virtual agent guide the customer through various store locations or windows to find the right resources and solutions for customer needs. With a visual spatial understanding of the entire physical environment, we expect our virtual agent to determine different strategies to help customers navigate through the space, introducing them to different areas or windows based on real-time location insights such as crowd levels and availability. This ensures an optimized strategy for addressing complex problems or needs, providing a seamless and efficient customer experience.

\textbf{\textit{Detailed Customer Experience Logging}} - Bridging digital and physical retail experience further offers design with regards to security, financial compliance, and auditing. Our virtual agent with visual understanding capability offers more transparency through its visual logs along with other sensor data captured by the mixed reality device. For example, if the virtual agent assumes the role of financial advisor, there is less threat of maleficent behavior and both the customer and banker are protected. Not only are transactions logged in this case, but our AI agent may also log visual and verbal records for additional analysis for sentiment and other indicators. 


\subsection{Customer Profile Memorization}
A common practice in financial and retail services is customer profiling. Banks and retail stores utilize different strategies to generate customer profiles to enhance quality of service and tailor services to meet specific needs of the individual. In our solution, we look to bridge a one-to-one customer profiling experience with real-time analysis so bank and store staff can proactively offer service recommendations tailored to the customer. This one-to-one profiling strategy also enables enhanced transactional security through anomaly and pattern detection, reducing the overall risk and threats to customer accounts and personal data.

\subsubsection{MR-based customer profiling for in-store and remote interaction} 
Our mixed reality workflow allows for a retail in-store virtual agent to be integrated into our existing online digital platform, providing a unique user experience. This bridge between online and retail in-store experience provides the opportunity for the system to draw on data sources such as: spending patterns, account history, savings plans, and other information. Introducing a mixed reality experience into the retail setting offers opportunity to mix the digital and human workflows. For example, when the customer engages with our virtual agent in a retail setting, we are able to offer direct customer service from a remote human staff using the same digital embodiment. However, since the user is in mixed reality, we envision a system that will enhance their experience with digital services while providing the necessary security required such as identity and application authorization workflows.

\subsection{Customer Experience Optimization and Operational Efficiency}
The system provides a mixed reality view of a virtual agent by default, reducing the need for initial human staff involvement. This approach provides a smoother automation workflow and possibly increases operation efficiency. In our solution, online banking as a system, enhances transparency for auditing with access to a comprehensive view of the customer's transactions, records, and importantly providing a transcription of spoken exchange. Our proposed architecture provides a direct interface to the users' data with real-time updates in a centrally hosted platform, which increases operation efficiency through automatic reporting and compliance to regulatory bodies. This removes the need for human transcribed reports and improves accuracy using consistent reporting frameworks. Our virtual agent is more than a user interface component; it works across various aspects of the core financial infrastructure to support the decision making process. 

\subsubsection{Distributed Agent Placement}
As the agent is physically located, each instance has a unique location and therefore a unique field of view and perspective of the retail space. The agent has a camera as part of its structure that is pointed at a designated area in the retail space. The camera may be fixed or rotational as to mimic how a human sees the area around them. As humans, our field of view alerts us to what is important in a given space \cite{HASSAN20072115}. We think of this setup as being analogous to a line of human bank tellers or cashiers you see when entering a branch. In our case, each cashier window is equipped with a camera, microphone, speaker, location beacon, among other sensors. Having a physical location, along with networking capabilities such as viewing the space through another agent's location, allows the agents to function as an independent service provider when a human staff is absent, or collaborate with human staff by their space management capabilities. 

\subsubsection{Proactive Interaction Initiation}
We designed the agent with multi-modal sensor input for initiating user interaction. In this way, the agent may engage in dialog with the user independently, as opposed to a user initiating the interaction. In these scenarios, the configuration of each agent's field of view opens new research opportunities for understanding how to manage event priority through response handling and delegation. For example, depending on the configuration and layout of the system, customers can be serviced simultaneously or redirected to additional agents based on their service needs, potentially opening up new opportunities in queue management and branch layout optimization. The operational efficiency of the system further scales without a proportional increase in cost, allowing the branch to increase workloads effectively.


\subsection{End User Privacy and Security}
The system decouples the agent, its digital representation (a 3D avatar), and inference engine (e.g., GPU processing server) into separate applications. While the avatar application runs on the end-user’s device as the main rendering and interaction entry point, the agent application runs on hardware that is physically located in the branch. The avatar application provides a customized user interface displaying both visual and text information workflows. The agent application provides the location, situational awareness, and queue management necessary for detecting when customers are within its range and where to guide them for optimal assistance. Splitting the avatar and agent into separate applications not only improves the functionality but also enhances end-user privacy and security. Compared to a conventional all-in-one kiosk solution commonly used in banking and retail environments, our method of deploying separate applications has the following benefits to potentially address common security and privacy issues: 

\textbf{\textit{Shoulder Surfing and Bystander Peeking}} - With the avatar projected in mixed reality on the end-user's device, we can adopt the same type of bystander peeking protection techniques we perform when using our mobile devices. If the user's MR device is running on a mobile phone, they can physically adjust their bodies or walk away to another location. At the same time, as our virtual agent has its own visual awareness, the agent might provide reminders to the user of their surroundings and possible incidents of such privacy violation. For HMD-based mixed reality or AR glasses, the customers will have more exclusive and private views of their information, further lowering the risk of bystander peeking.

\textbf{\textit{Authentication and Authorization}} - We can establish the user session with standard authentication and authorization using biometric or knowledge-based authentication solutions present on the end user's headset or mobile device. As customer profiles and financial data may contain sensitive information, the system can build on techniques for managing data through user entitlements and roles. We are building the system following the principle of least privilege, an information security concept that limits access to the minimum amount of permissions or access needed to perform the task at hand \cite{doi:10.1080/08874417.2022.2128937}.

\textbf{\textit{Location-based Session Management}} - Establishing a new session per user may require sophisticated infrastructural logic to understand when users have physically changed places in line or in the store. Further, running the agent and avatar applications on their own devices allows us to address these issues more easily, as our agent application becomes less dependent on end-user data for executing its activities and responsibilities.

\section{Architecture}
This section outlines our current implementation and system architecture for a banking usage scenario. 
\subsection{System Overview}
The system is developed around three foundational components: 1) a centralized banking service for customer related data communication and VLM model inference, 2) one or more physically located multi-modal sensors representing the virtual agent's physical embodiment, and 3) one or more end-user client applications providing a personalized user experience. The centralized banking service can be hosted in the cloud, within a bank's data center, or on a hybrid cloud platform. The client applications currently run on mobile devices such as iPhones, iPads, or mixed reality headsets. These client applications interface with the centralized inference engine and banking service through a client-server architecture. 

\begin{figure}[hbt!]
    \centering
    \includegraphics[width=1\linewidth]{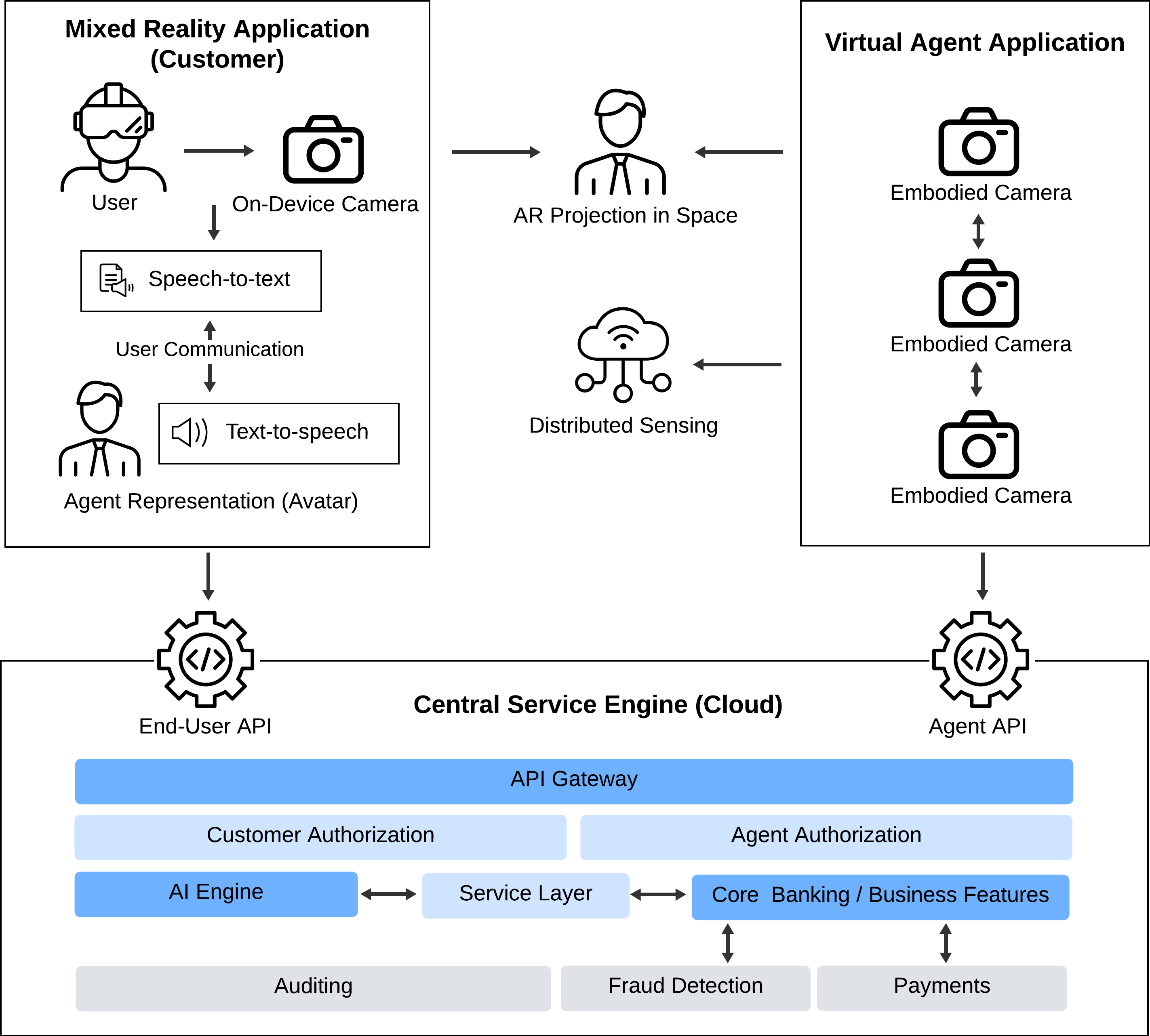}
    \caption{System level block diagram. Our system consists of: 1) a MR customer application for interacting with the agent. 2) A virtual agent application embodied in multiple cameras in the physical environment. 3) a central service engine on the cloud for both agent and end-user authentication, AI inference, and general services.}
    \label{fig:system}
\end{figure}

An early goal of the project was to give the agent an independent field of view. We did not want the agent's camera to be tied to the end-user application, instead we wanted it to have its own unique perspective. To achieve this, we created two independent applications running on separate devices. We were pleased to discover splitting the applications revealed many additional advantages. For example, decoupling the end-user application from the physically based agent simplified the agent code by offloading security and other functionality to the end-user's device. Additionally, we had the opportunity to personalize the experience for the user. For example, translating the dialog into a preferred language and localized context and providing the necessary accessibility required. A block diagram showing the logical components is depicted in Figure \ref{fig:system}.

\subsection{System Component Details}
Our system is designed around a cloud-based AI banking back-end for mobile and IoT based applications. The cloud based system provides a scalable infrastructure to support many clients simultaneously. The architecture may automatically scale up or down our inference service, depending on user load, while providing additional interfaces to core banking, data services, payments, auditing, and other financial services. Below we detail the following components:

\textbf{\textit{Agent Application}} - The agent application is deployed with one or more physical embodiments installed throughout a simulated retail space. These agents are mounted on stationary tripods with a fixed field of view, representing the agents' eyes. The application controlling the agents runs on an iPad, featuring a minimal UI used to setup the interface with the camera and AI inference service. The agent application uses Apple's comprehensive AVFoundation framework \cite{apple-avfoundation} for working with audiovisual media. The framework allows for media capture in the form of buffer capture, capturing video frames at 480 x 360 resolution. The application converts each frame to PNG format prior to sending it to the server. 

\textbf{\textit{Avatar Application}} The second client application provides the user experience. When the user approaches one of the agents in the space the system will proactively engage them through beacon ranging. Apple's Core Location \cite{apple-corelocation} framework provides location and heading information for ranging the devices using Bluetooth low-energy (BLE). When the user's application is notified it has come within range of the physically located agent application, the application opens a WebSocket in the background to the central banking service. The banking service is then free to engage the AI agent and establish a dialog. We find this to be an effective natural interaction similar to a human-to-human banking scenario where the bank tellers almost always control the customer queue by directing the client when it is their turn to be served.  

\textbf{\textit{Centralized Banking Service}} - The server architecture implements an AI inference micro-service for understanding text and vision. Specifically, we have tested our prototype with Microsoft's Phi-3-vision-128K-instruct model \cite{abdin2024phi}. This model is targeted for single image input combined with text prompts. As a prototype, to understand the parameters around text and vision understanding, we implemented inference as stateless service, where an end-user or the agent may establish the banking session. The model is an English language based model so any language translation needs to occur before or after the inference. 

\begin{figure}[hbt!]
    \centering
    \includegraphics[width=1\linewidth, trim={0 30 0 28},clip]{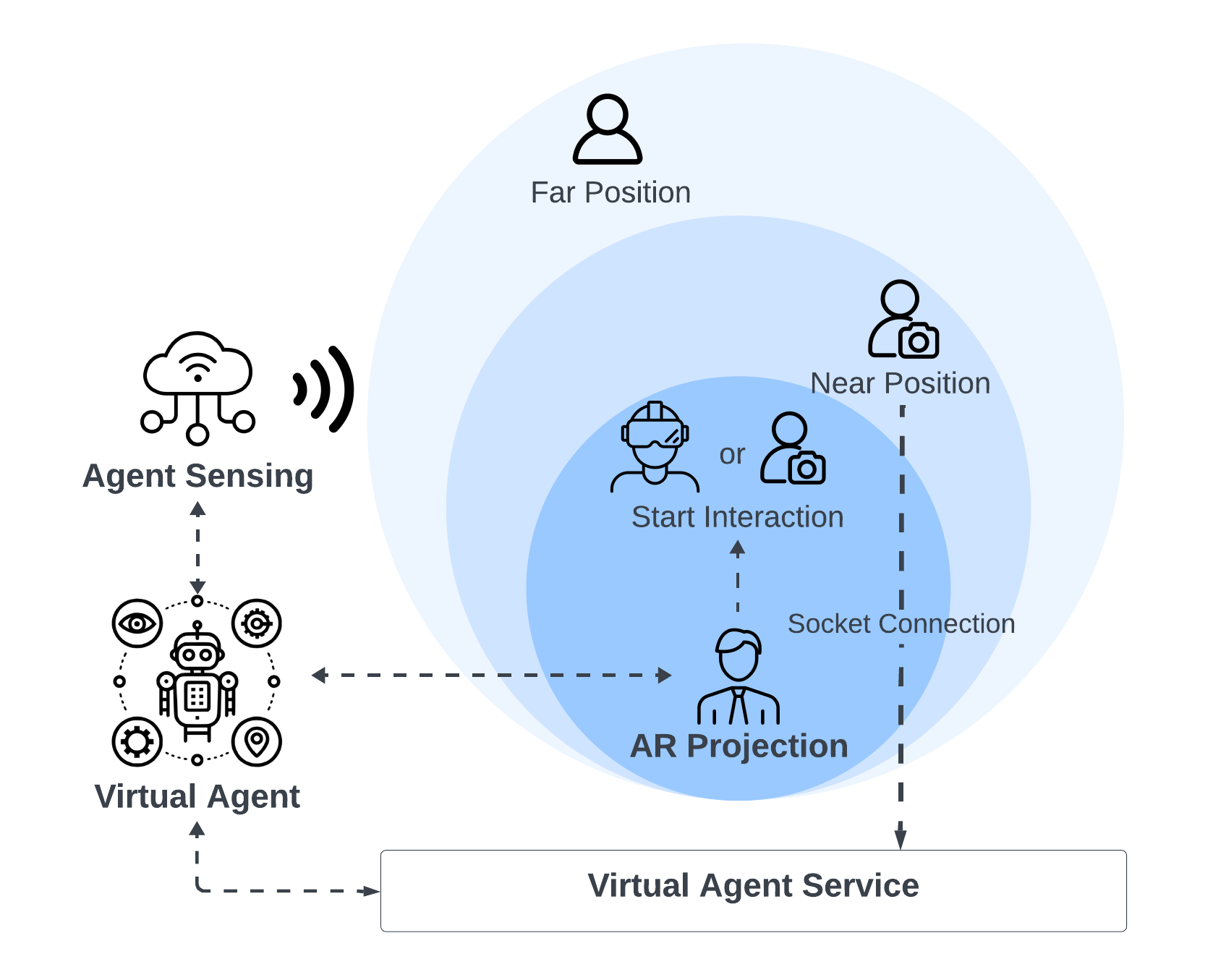}
    \caption{An overview of our Avatar/Agent beacon ranging based on customer position and distance to service areas.}
    \label{fig:beaconranging}
\end{figure}

\textbf{\textit{IoT-based Service Launcher}}
Integrating the vision language model into agent/avatar architecture exposed challenges creating a more human-like fluid experience. To address this, we adopted IoT sensors for understanding the location and position of the end-user relative to the agent. With this integration we can enhance customer queue processing. As described in the avatar application, Core Location provides beacon ranging and we implement this sensing feature as a background service of the app so that the user does not need to proactively open their banking application for discovery by the agent. This feature allows us to reduce connection latency by establishing the connection early as the customer moves toward the kiosk. Once the user has been engaged by the agent, they can open their banking application and establish a working session. Figure \ref{fig:beaconranging} depicts the scenario where an end-user approaches an agent in a confined space, as the user approaches the signal increases and the range intent sends an event to the both the end-user and agent applications. 

\section{Discussion}
While our current system focuses on facilitating context-aware agent interactions for customers and enabling proactive responses to their needs, we are also actively researching solutions to areas such as minimizing the effect of hallucination and supporting selective memorization of user input. Common in many AI applications these days, a VLM backed by an LLM might generate false or irrelevant information. This issue is particularly critical in a customer service environment where precision and relevance are key to maintaining customer trust. The more responsibilities an AI agent is taking, the more crucial it is to avoid hallucination. To mitigate such risk, we are exploring methods to enhance the overall performance of VLM and improve validation. 

Another area of focus is the privacy and management of customer data. Given the sensitive nature of the information handled by our system, it is imperative to develop and implement strategies for selective memorization. A banking or retail environment can be highly dynamic with challenging needs from the customers every day. For our virtual agent, it is critical for the system to ensure adherence to financial regulatory compliance. The system must be transparent in its data usage while ensuring data retention policies are in place. Investigating the feasibility of audit through periodically refreshing the agent infrastructure to address data concerns and privacy protections will not only increase customer confidence but improve regulator trust in our system.

These enhancements are crucial not only for aligning with stringent data protection standards but also for elevating the overall effectiveness and ethical standing of our virtual agent. By addressing these challenges, we aim to refine out system's capabilities, ensuring it meets both user and legal expectations in providing a secure, reliable and customer-centric service in financial environments.



\section{Future Work and Conclusion}
In this work, we present a novel approach to enhancing customer service in financial and retail environments through the development of a context-aware 3D virtual agent integrated with mixed reality. Our system leverages vision language models to provide dynamic and intelligent situation understanding to meet the unique needs of each customer in a dynamic and complex environment. We have outlined key design considerations focusing on proactive and personalized support, situational awareness, and privacy assurance. Future work will aim to refine aspects such as hallucination and data privacy management to ensure our system not only supports but also enriches the customer experience with high levels of security, compliance, and personalization.

\section{Disclaimer}

This paper was prepared for informational purposes by the technology research division of JPMorgan Chase. This paper is not a product of the Research Department of this institution or its affiliates. Neither the institution nor any of its affiliates makes any explicit or implied representation or warranty and none of them accept any liability in connection with this paper, including, without limitation, with respect to the completeness, accuracy, or reliability of the information contained herein and the potential legal, compliance, tax, or accounting effects thereof. This document is not intended as investment research or investment advice, or as a recommendation, offer, or solicitation for the purchase or sale of any security, financial instrument, financial product or service, or to be used in any way for evaluating the merits of participating in any transaction.

\bibliographystyle{abbrv-doi}

\bibliography{template}

\begin{thebibliography}{10}

\bibitem{abdin2024phi}
M.~Abdin, S.~A. Jacobs, A.~A. Awan, J.~Aneja, A.~Awadallah, H.~Awadalla, N.~Bach, A.~Bahree, A.~Bakhtiari, H.~Behl, et~al.
\newblock Phi-3 technical report: A highly capable language model locally on your phone.
\newblock {\em arXiv preprint arXiv:2404.14219}, 2024.

\bibitem{apple-avfoundation}
Apple audio and video avfoundation is the full featured framework for working with time-based audiovisual media on ios, macos, watchos and tvos, 2024.

\bibitem{apple-corelocation}
Apple core location provides services that determine a device’s geographic location, altitude, and orientation, or its position relative to a nearby ibeacon device., 2024.

\bibitem{bailenson2004}
J.~N. Bailenson and J.~Blascovich.
\newblock Avatars.
\newblock {\em Encyclopedia of human-computer interaction}, 1:64--6, 2004.

\bibitem{bonetti2018}
F.~Bonetti, G.~Warnaby, and L.~Quinn.
\newblock {\em Augmented Reality and Virtual Reality in Physical and Online Retailing: A Review, Synthesis and Research Agenda}, pp. 119--132.
\newblock Springer International Publishing, Cham, 2018. doi: {{%
10\hspace{.1pt}\discretionary{.}{%
}{.}\hspace{.4pt}1007\discretionary{/}{%
}{/}978\discretionary{%
}{-}{-}3\discretionary{%
}{-}{-}319\discretionary{%
}{-}{-}64027\discretionary{%
}{-}{-}3\_9}}


\bibitem{bordes2024introduction}
F.~Bordes, R.~Y. Pang, A.~Ajay, A.~C. Li, A.~Bardes, S.~Petryk, O.~Ma{\~n}as, Z.~Lin, A.~Mahmoud, B.~Jayaraman, et~al.
\newblock An introduction to vision-language modeling.
\newblock {\em arXiv preprint arXiv:2405.17247}, 2024.

\bibitem{chen2022pali}
X.~Chen and X.~Wang.
\newblock Pali: Scaling language-image learning in 100+ languages.
\newblock In {\em Conference on Neural Information Processing Systems (NeurIPS)}, 2022.

\bibitem{cherakara2023}
N.~Cherakara, F.~Varghese, S.~Shabana, N.~Nelson, A.~Karukayil, R.~Kulothungan, M.~A. Farhan, B.~Nesset, M.~Moujahid, T.~Dinkar, V.~Rieser, and O.~Lemon.
\newblock Furchat: An embodied conversational agent using llms, combining open and closed-domain dialogue with facial expressions, 2023.

\bibitem{crolic2021}
C.~Crolic, F.~Thomaz, R.~Hadi, and A.~T. Stephen.
\newblock Blame the bot: Anthropomorphism and anger in customer–chatbot interactions.
\newblock {\em Journal of Marketing}, 86:132 -- 148, 2021.

\bibitem{das2023}
S.~Das, C.~Faklaris, J.~I. Hong, and L.~A. Dabbish.
\newblock {\em The Security \& Privacy Acceptance Framework (SPAF)}.
\newblock 2023.

\bibitem{guo2023mobile}
S.~Guo and G.~Tocquer.
\newblock Customer experience and mobile application design.
\newblock In {\em Proceedings of the 2023 3rd International Conference on Human Machine Interaction}, ICHMI '23, p. 58–64. Association for Computing Machinery, New York, NY, USA, 2023.

\bibitem{guo2024largelanguagemodelbased}
T.~Guo, X.~Chen, Y.~Wang, R.~Chang, S.~Pei, N.~V. Chawla, O.~Wiest, and X.~Zhang.
\newblock Large language model based multi-agents: A survey of progress and challenges, 2024.

\bibitem{HASSAN20072115}
S.~E. Hassan, J.~C. Hicks, H.~Lei, and K.~A. Turano.
\newblock What is the minimum field of view required for efficient navigation?
\newblock {\em Vision Research}, 47(16):2115--2123, 2007. doi: {{%
10\hspace{.1pt}\discretionary{.}{%
}{.}\hspace{.4pt}1016\discretionary{/}{%
}{/}j\hspace{.1pt}\discretionary{.}{%
}{.}\hspace{.4pt}visres\hspace{.1pt}\discretionary{.}{%
}{.}\hspace{.4pt}2007\hspace{.1pt}\discretionary{.}{%
}{.}\hspace{.4pt}03\hspace{.1pt}\discretionary{.}{%
}{.}\hspace{.4pt}012}}


\bibitem{jabbar2019}
J.~Jabbar, I.~Urooj, W.~JunSheng, and N.~Azeem.
\newblock Real-time sentiment analysis on e-commerce application.
\newblock In {\em 2019 IEEE 16th International Conference on Networking, Sensing and Control (ICNSC)}, pp. 391--396, 2019.

\bibitem{konenkov2024vrgptvisuallanguagemodel}
M.~Konenkov, A.~Lykov, D.~Trinitatova, and D.~Tsetserukou.
\newblock Vr-gpt: Visual language model for intelligent virtual reality applications, 2024.

\bibitem{Lao2021ExploringHD}
A.~Lao, M.~Vlad, and A.~Mart{\'i}n.
\newblock Exploring how digital kiosk customer experience enhances shopping value, self-mental imagery and behavioral responses.
\newblock {\em International Journal of Retail \& Distribution Management}, 2021.

\bibitem{li2019acousticsentiment}
B.~Li, D.~Dimitriadis, and A.~Stolcke.
\newblock Acoustic and lexical sentiment analysis for customer service calls.
\newblock In {\em ICASSP 2019 - 2019 IEEE International Conference on Acoustics, Speech and Signal Processing (ICASSP)}, pp. 5876--5880, 2019. doi: {{%
10\hspace{.1pt}\discretionary{.}{%
}{.}\hspace{.4pt}1109\discretionary{/}{%
}{/}ICASSP\hspace{.1pt}\discretionary{.}{%
}{.}\hspace{.4pt}2019\hspace{.1pt}\discretionary{.}{%
}{.}\hspace{.4pt}8683679}}


\bibitem{li2022mplug}
C.~Li, H.~Xu, J.~Tian, W.~Wang, M.~Yan, B.~Bi, J.~Ye, H.~Chen, G.~Xu, Z.~Cao, et~al.
\newblock mplug: Effective and efficient vision-language learning by cross-modal skip-connections.
\newblock {\em arXiv:2205.12005}, 2022.

\bibitem{liu2023instructionfollowing}
H.~Liu, L.~Lee, K.~Lee, and P.~Abbeel.
\newblock Instruction-following agents with jointly pre-trained vision-language models, 2023.

\bibitem{Nagao1998AgentAR}
K.~Nagao.
\newblock Agent augmented reality : Agents integrate the real world with cyberspace.
\newblock 1998.

\bibitem{niu2024screenagentvisionlanguagemodeldriven}
R.~Niu, J.~Li, S.~Wang, Y.~Fu, X.~Hu, X.~Leng, H.~Kong, Y.~Chang, and Q.~Wang.
\newblock Screenagent: A vision language model-driven computer control agent, 2024.

\bibitem{obermeier2020}
G.~Obermeier, R.~Zimmermann, and A.~Auinger.
\newblock The effect of queuing technology on customer experience in physical retail environments.
\newblock In F.~F.-H. Nah and K.~Siau, eds., {\em HCI in Business, Government and Organizations}, pp. 141--157, 2020.

\bibitem{doi:10.1080/08874417.2022.2128937}
M.~Plachkinova and K.~Knapp.
\newblock Least privilege across people, process, and technology: Endpoint security framework.
\newblock {\em Journal of Computer Information Systems}, 63(5):1153--1165, 2023.

\bibitem{schobel2023}
S.~Schöbel, A.~Schmitt, D.~Benner, M.~Saqr, A.~Janson, and J.~M. Leimeister.
\newblock Charting the evolution and future of conversational agents: A research agenda along five waves and new frontiers.
\newblock {\em Information Systems Frontiers}, 26:1--26, 04 2023.

\bibitem{tlapana2021}
T.~Tlapana.
\newblock The impact of store layout on consumer buying behaviour: A case of convenience stores from a selected township in kwazulu natal.
\newblock {\em International Review of Management and Marketing}, 11(5):1–6, Sep. 2021.

\bibitem{wan2024}
H.~Wan, J.~Zhang, A.~A. Suria, B.~Yao, D.~Wang, Y.~Coady, and M.~Prpa.
\newblock Building llm-based ai agents in social virtual reality.
\newblock In {\em Extended Abstracts of the 2024 CHI Conference on Human Factors in Computing Systems}, CHI EA '24. Association for Computing Machinery, New York, NY, USA, 2024. doi: {{%
10\hspace{.1pt}\discretionary{.}{%
}{.}\hspace{.4pt}1145\discretionary{/}{%
}{/}3613905\hspace{.1pt}\discretionary{.}{%
}{.}\hspace{.4pt}3651026}}


\bibitem{wang2019}
I.~Wang, J.~Smith, and J.~Ruiz.
\newblock Exploring virtual agents for augmented reality.
\newblock In {\em Proceedings of the 2019 CHI Conference on Human Factors in Computing Systems}, CHI '19, p. 1–12. Association for Computing Machinery, New York, NY, USA, 2019.

\bibitem{wang2022git}
J.~Wang, Z.~Yang, X.~Hu, L.~Li, K.~Lin, Z.~Gan, Z.~Liu, C.~Liu, and L.~Wang.
\newblock Git: A generative image-to-text transformer for vision and language.
\newblock {\em arXiv:2205.14100}, 2022.

\bibitem{beit3}
W.~Wang, H.~Bao, L.~Dong, J.~Bjorck, Z.~Peng, Q.~Liu, K.~Aggarwal, O.~K. Mohammed, S.~Singhal, S.~Som, and F.~Wei.
\newblock Image as a foreign language: {BEiT} pretraining for vision and vision-language tasks.
\newblock In {\em Proceedings of the IEEE/CVF Conference on Computer Vision and Pattern Recognition}, 2023.

\bibitem{wieland2024}
D.~A. Wieland, B.~S. Ivens, E.~Kutschma, and P.~A. Rauschnabel.
\newblock Augmented and virtual reality in managing b2b customer experiences.
\newblock {\em Industrial Marketing Management}, 119:193--205, 2024.

\bibitem{zhang2024buildingcooperativeembodiedagents}
H.~Zhang, W.~Du, J.~Shan, Q.~Zhou, Y.~Du, J.~B. Tenenbaum, T.~Shu, and C.~Gan.
\newblock Building cooperative embodied agents modularly with large language models, 2024.

\end{thebibliography}
\end{document}